\documentclass[aps,showpacs,twocolumn]{revtex4}
\usepackage{graphicx}
\usepackage{amssymb}
\begin{document}
\title{Spin-polarized transport in a lateral two-dimensional diluted
magnetic semiconductor electron gas}
\author{W. Yang}
\author{Kai Chang}
\altaffiliation[Author to whom correspondence should be addressed. Electronic address:]{kchang@red.semi.ac.cn.}
\author{X. G. Wu}
\author{H. Z. Zheng}
\affiliation{NLSM, Institute of Semiconductors, Chinese Academy of Sciences, P. O. Box
912, Beijing 100083, China}

\begin{abstract}
The transport property of a lateral two-dimensional diluted magnetic
semiconductor electron gas under a spatially periodic magnetic field is
investigated theoretically. We find that the electron Fermi velocity along
the modulation direction is highly spin-dependent even if the spin
polarization of the carrier population is negligibly small. It turns out
that this spin-polarized Fermi velocity alone can lead to a strong spin
polarization of the current, which is still robust against the energy
broadening effect induced by the impurity scattering.
\end{abstract}

\maketitle

The spin of carriers in semiconductors has attracted considerable attention
recently due to its great importance in the development of spintronic devices%
\cite{SPT} and quantum computation\cite{QC}. One of the main obstacles for
these potential applications is to inject highly spin-polarized carriers
into semiconductors. The spin-polarized current through a diluted magnetic
semiconductor (DMS)\ junction was predicted theoretically\cite{Egues} and
demonstrated experimentally by applying a strong magnetic field at low
temperatures\cite{DMSTUN}. The spin polarization of the current defined as $%
P_{J}=(J_{\uparrow }-J_{\downarrow })$ /$(J_{\uparrow }+J_{\downarrow })$ in
semiconductors comes from two physical mechanisms. One is the contribution
from the spin polarization of carrier population $P_{c}=(n_{e}^{\uparrow
}-n_{e}^{\downarrow })/(n_{e}^{\uparrow }+n_{e}^{\downarrow })$. Another
important contribution is due to the difference in the spin-dependent Fermi
velocity of carriers $v_{F}^{\uparrow ,\downarrow }$, since the electric
current depends also on the electron group velocity at the Fermi energy $%
J_{\uparrow ,\downarrow }=qn_{e}^{\uparrow ,\downarrow
}(E_{F})v_{F}^{\uparrow ,\downarrow }$ ($q$ is the carrier charge). It
should be emphasized that the Fermi velocity $v_{F}^{\uparrow ,\downarrow }$
can be engineered utilizing the spin-dependent exchange interaction, the
geometric dimensionality and even the tailoring of the sample structure.
Most of the previous studies focus on seeking the spin polarized current
induced by the high spin polarization of carriers.

In this work we propose a structure of DMS lateral superlattice (LSL) \cite%
{LDEG,CMHU} and demonstrate theoretically that a strongly spin-polarized
current can be generated even when the carrier spin polarization is
vanishingly small. We show that in such a weakly spin-polarized system, the
strong spin polarization of the current stems mainly from the difference
between spin-up and spin-down Fermi velocities, instead of the spin
polarization of the carriers.

\begin{figure}[tbp]
\includegraphics[width=\columnwidth]{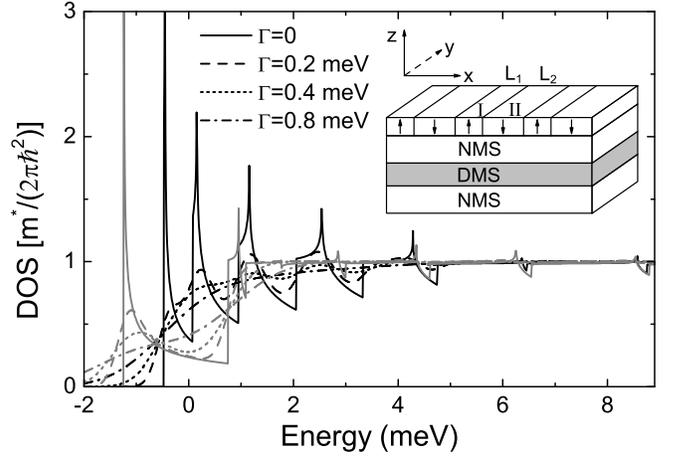}
\caption{The spin-up (black lines) and spin-down (gray lines) DOS for
different energy broadenings $\Gamma $=0 (solid lines), 0.2 meV (dashed
lines), 0.4 meV (dotted lines), and 0.8 meV (dash-dotted lines). The inset
shows schematically the DMS 2DEG sandwiched between NMS layers, with
magnetic stripes deposited above.}
\end{figure}
As schematically shown in the inset of Fig. 1, the DMS\ layer is sandwiched
by two nonmagnetic semiconductor (NMS) layers. Magnetic stripes with
alternating magnetization directions and widths ($L_{1}$ and $L_{2},$
respectively) are deposited periodically on the top of the NMS\ layer. The
periodic magnetic field generated by the magnetic stripes penetrates into
the DMS LSL and magnetizes the Mn ions, which lead to spin splitting of the
conduction electron, due to their \textit{s-d} exchange interaction.

The Hamiltonian for the two-dimensional electron gas in the DMS LSL reads 
\begin{equation}
H=\frac{(\mathbf{p}+e\mathbf{A})^{2}}{2m^{\ast }}+H_{\text{ex}},
\label{Hamil}
\end{equation}%
here $m^{\ast }$ is the effective mass, $\mathbf{A}=[0,A(x),0]$ is the
vector potential for the periodic magnetic field $\mathbf{B}(x)=B(x)\mathbf{e%
}_{z},$ where $B(x)=B(x+L)$ has a constant value $B_{1}$ ($B_{2}$) in region
I (II). The average magnetic field $B_{\text{av}}\equiv
(B_{1}L_{1}+B_{2}L_{2})/L$ should be zero due to the sourceless nature of
the magnetic field.\cite{Ibrahim,Shijirong} This feature ensures that the
Hamiltonian is periodic along the $x$ axis.%
\begin{equation}
H_{\text{ex}}=-\sum\limits_{i}J_{s-d}(\mathbf{r}-\mathbf{R}_{i})\mathbf{s}%
\cdot \mathbf{S}_{i}  \label{Hex}
\end{equation}%
describes the \textit{s-d} exchange interaction between conduction electrons
and the 3d$^{5}$ electrons of Mn ions, where $J_{^{s-d}}$ is the \textit{s-d}
exchange integral, $\mathbf{r}$ ($\mathbf{R}_{i}$) and $\mathbf{s}$ (\textbf{%
S}$_{i}$) are the coordinate and spin of the band electron (Mn ions),
respectively. The summation runs over the lattice sites occupied by the Mn
ions. Within the mean field approximation, we have $H_{\text{ex}}=\sigma
_{z}\Delta (x)/2$, where $\sigma _{z}$ is the $z$ component of the Pauli
matrices and $\Delta (x)=N_{0}\alpha xS_{0}B_{5/2}\left\{ g_{Mn}S\mu _{B}B/%
\left[ k_{B}(T+T_{0})\right] \right\} $, with $B_{J}(x)$ the Brillouin
function, $S=5/2$ the 3d$^{5}$ spin of the Mn ion, $N_{0}$ the number of
cation sites per unit volumn, $\alpha $ the \textit{s-d }exchang constant, $%
S_{0}$ and $T_{0}$ phenomenological parameters describing the
antiferromagnetic superexchange between neighboring Mn ions.

Since $p_{y}$ is a constant of motion, the electron Hamiltonian can be
reduced to one-dimension effective Hamiltonian $H_{1D}=p_{x}^{2}/(2m^{\ast
})+V_{\text{eff}}(x),$ characterized by the periodic spin-dependent potential%
\[
V_{\text{eff}}(x)=V_{\text{eff}}(x+L)=\frac{1}{2m^{\ast }}[\hbar
k_{y}+eA(x)]^{2}+\sigma _{z}\Delta _{z}(x)/2. 
\]%
Using a plane wave basis, the eigenvalues $E_{n\sigma }(\mathbf{k})$ and
eigenstates $\psi _{n\mathbf{k}\sigma }(\mathbf{r})$ can be obtained
numerically, where $\sigma =\uparrow ,\downarrow $ denotes the spin
components. From the space inversion invariance of the Hamiltonian, we see
that $E_{n\sigma }(\mathbf{k})$ is invariant under the operation $%
k_{x}\rightarrow -k_{x}$ or $k_{y}\rightarrow -k_{y}$.

The scattering effect of impurities (including the remaining parts of the 
\textit{s-d} exchange interaction not included in the mean-field
approximation) can be modelled by introducing a finite broadening $\Gamma
=\hbar /\tau _{c}$ to the energy levels, where $\tau _{c}$ is the carrier
lifetime. As a result, the spin-dependent DOS per unit area is given by 
\begin{equation}
D_{\sigma }(E)=\sum_{n}\int \frac{d^{2}\mathbf{k}}{(2\pi )^{2}}P\left[
E-E_{n\sigma }(\mathbf{k})\right] ,  \label{DOS}
\end{equation}%
where $P(E)=1/(\sqrt{2\pi }\Gamma )\exp \left[ -E^{2}/(2\Gamma ^{2})\right] $
is the Gaussian broadening function. The spin-dependent conductivity tensor
at low temperature is given by $(i,j=x,y)$

\begin{equation}
\sigma _{ij}^{\sigma }=e^{2}\tau \sum_{n}\int \frac{d^{2}\mathbf{k}}{(2\pi
)^{2}}v_{n\sigma }^{i}(\mathbf{k})v_{n\sigma }^{j}(\mathbf{k})P\left[
E_{F}-E_{n\sigma }(\mathbf{k})\right] ,
\end{equation}%
where $\mathbf{v}_{n\sigma }(\mathbf{k})=(1/\hbar )\nabla _{\mathbf{k}%
}E_{n\sigma }(\mathbf{k})$ is the group velocity and $\tau $ is the
transport relaxation time. Since the energy eigenvalue $E_{n\sigma }(\mathbf{%
k})$ is an even function of $k_{x}$ or $k_{y}$, all the off-diagonal
conductivities vanish.

We use the following parameters for II-VI DMS Zn$_{1-x}$Mn$_{x}$Se in our
calculation: $m^{\ast }=0.13m_{0}$ ($m_{0}$\ is the free electron mass), $%
g_{Mn}=2.0$, $N_{0}\alpha =270$ meV, $x=0.11$, $S_{0}=0.91$, $T_{0}=1.4$ K,%
\cite{Fatah,para, Lee, Jeon} $T=1$ K, $B_{1}=0.3$ T, $L_{1}=L_{2}/4=26$ nm.
The energy broadening parameter $\Gamma $ is determined by the quality of
the DMS 2DEG. It can be roughly estimated from the experimentally measured
transport relaxation time $\tau $ by assuming $\tau _{c}=\tau $ (this
relation holds exactly for impurity scattering by central symmetric
scatterers). For non-DMS materials such as GaAs/AlGaAs 2DEG, the electron
mobility can reach several tens of m$^{2}/($V s$)$, while for DMS 2DEG, the
mobility is decreased due to the scattering of the magnetic ions. However,
high mobility II-VI DMS 2DEG has already been realized, e.g., in HgMnTe [$%
\mu $=$5$ m$^{2}/($V s$)$] and CdMnTe [$\mu $=$6$ m$^{2}/($V s$)$].\cite%
{Grabecki,Teran} Correspondingly, the energy broadening for ZnMnSe with
mobility of the order of several m$^{2}/($V s$)$ should be smaller than 1.0
meV. From our numerical results (to be discussed below), the energy
broadening factor $\Gamma $ is not so crucial for the spin polarization of
the current.

Due to the spin-dependent effective potential $V_{\text{eff}}(x)$, the spin
degeneracy of the electron is lifted, resulting in the spin-dependent DOS
shown in Figs. 1. For vanishing energy broadening, the spin-dependent DOS
shows sharp peaks (van Hove singularities) at low energy, which come from
the one dimensional nature of the electron confined by $V_{\text{eff}}(x)$.
At higher energies, the strengths of the peaks decrease, and the DOS
approaches the value\ of a free 2DEG $D_{0}=m^{\ast }/(2\pi \hbar ^{2})$,
since the influence of $V_{\text{eff}}(x)$ decreases. When finite energy
broadening is taken into account, the van Hove singularities are smoothed
out, even for the smallest energy broadening $\Gamma =0.2$ meV.

\begin{figure}[tbp]
\includegraphics[width=\columnwidth]{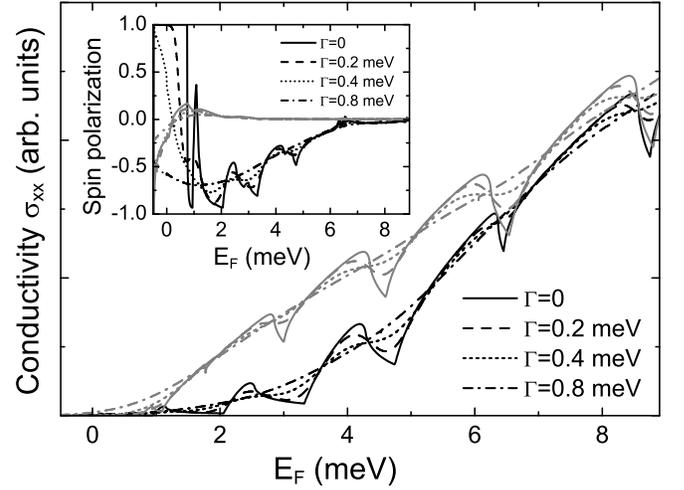}
\caption{Spin-up (black lines) and spin-down (gray lines) conductivities $%
\protect\sigma _{\text{xx}}$ as functions of Fermi energy. Inset shows the
current polarization $P_{J}$ along the $x$ axis (black lines) and the
carrier polarization $P_{c}$ (gray lines). They are obtained from $\Gamma =$
0 (solid lines), 0.2 meV (dashed lines), 0.4 meV (dotted lines), and 0.8 meV
(dash-dotted lines).}
\end{figure}
Figure 2 shows the spin-dependent conductivities $\sigma _{\text{xx}}$ as
functions of Fermi energy $E_{F}$. We see that for vanishing energy
broadening, $\sigma _{\text{xx}}$ exhibits many peaks and dips, due to the
van Hove singularities in Fig. 1. With increasing energy broadening,
however, these fine structures are smoothed out. The most important feature,
as can be seen from the inset of Fig. 2, is the big difference between the
carrier polarization $P_{c}$ and current polarization $P_{J}$. The former is
apperciable at low $E_{F}$, but it vanishes rapidly with increasing $E_{F}$.
On the other hand, large $P_{J}$ persists up to a much higher Fermi energy.
In the region where $P_{c}$ vanishes, this large current polarization stems
mainly from the different Fermi velocities between electrons with different
spins, instead of the polarization of the carrier itself. Further, the
magnitude of $P_{J}$ is nearly not affected by the increasing energy
broadening, although the fine structures have been smoothed out. These
interesting features demonstrate that the Fermi velocity polarization alone
can also be utilized to generate a strong spin-polarized current in a weakly
polarized carrier system.

\begin{figure}[tbp]
\includegraphics[width=\columnwidth]{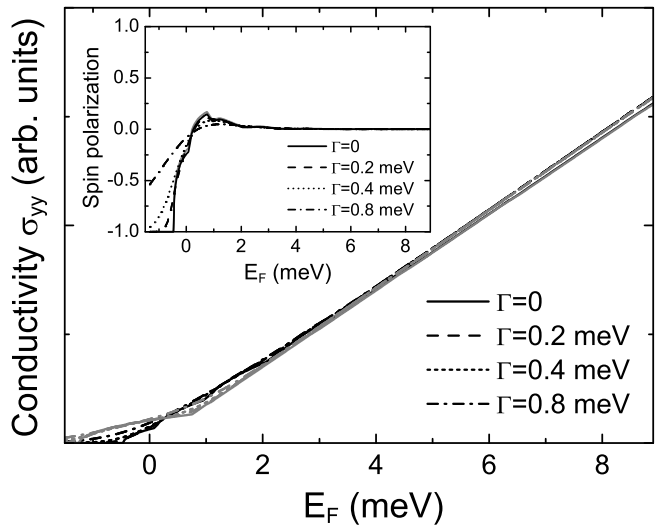}
\caption{Spin-up (black lines) and spin-down (gray lines) conductivities $%
\protect\sigma _{\text{yy}}$ as functions of Fermi energy. Inset: current
polarization along the $y$ axis (black lines) and carrier polarization (gray
lines) as functions of Fermi energy. They are obtained from $\Gamma $=0
(solid lines), 0.2 meV (dashed lines), 0.4 meV (dotted lines), and 0.8 meV
(dash-dotted lines).}
\end{figure}
Another interesting feature of this spin-polarized current is its anisotropy
arising from the anisotropic structure of the DMS LSL [see the inset of Fig.
1]. In Fig. 3, we plot the spin-dependent conductivities $\sigma _{\text{yy}%
} $ as functions of Fermi energy. It can be seen that $\sigma _{\text{yy}}$
increases monotonically with increasing Fermi energy. Further, the spin
polarization of $\sigma _{\text{yy}}$ (see the inset of Fig. 3) follows
closely the carrier polarization $P_{c}$. As a result, the current
polarization along the $y$ axis is always very small, except for very small
Fermi energy $E_{F}\lesssim 0$. These features are quite different from
those for $\sigma _{\text{xx}}$, due to the fact that the electron are
confined by $V_{\text{eff}}(x)$ along the $x$ axis, while it can move freely
along the $y$ axis. When the energy broadening is increased, the fine
structures of $\sigma _{\text{yy}}$ and its spin polarization are smoothed
out, similar to the case for $\sigma _{\text{xx}}$.

In summary, we have investigated theoretically the transport property of a
2DEG in a DMS LSL. Our numerical results demonstrate that a strongly
spin-polarized current can be generated utilizing the periodic magnetic
modulation in the DMS\ structure. The spin polarization in the carrier
population could be weak, but it is still possible to generate a strongly
spin-polarized current in the modulation direction. This spin polarization
of the current arises from the difference between spin-up and spin-down
Fermi velocities. It is still robust against the energy broadening effect
induced by the impurity scattering. Our theoretical work might provide a new
approach for exploiting planar spintronic devices, e.g., spin-polarized
electron sources and spin filters.

\begin{acknowledgments}
This work was partly supported by the NSFC No. 60376016 and the special fund
for Major State Basic Research Project No. G001CB3095 of China.
\end{acknowledgments}

\end{document}